\documentclass{article}

\usepackage{PRIMEarxiv}

\usepackage[utf8]{inputenc} 
\usepackage[T1]{fontenc}    
\usepackage{hyperref}       
\usepackage{url}            
\usepackage{booktabs}       
\usepackage{amsfonts}       
\usepackage{nicefrac}       
\usepackage{microtype}      
\usepackage{lipsum}
\usepackage{fancyhdr}       
\usepackage{graphicx}       
\graphicspath{{media/}}     
\usepackage{siunitx}
\usepackage{setspace}
\usepackage{amsmath}
\title{Magnetophoretic long jump of magnetic microparticles in an engineered magnetic stray field landscape for highly localized and large throughput on-chip fractionation
}

\author{
  Rico Huhnstock, Lukas Paetzold, Arno Ehresmann \\
  Institute of Physics and Center for Interdisciplinary Nanostructure Science and Technology (CINSaT) \\
  University of Kassel \\
  Heinrich-Plett-Str. 40, D-34132 Kassel, Germany\\
  \texttt{rico.huhnstock@physik.uni-kassel.de} \\
   \And
  Piotr Ku$\mathrm{\acute{s}}$wik \\
  Institute of Molecular Physics \\
  Polish Academy of Sciences \\
  M. Smoluchowskiego 17, Poznań, 60-179, Poland \\
}

\begin{document}
\maketitle
\onehalfspacing

\begin{abstract}
A common issue faced by magnetic particle-based Lab-on-a-chip systems, e.g, for medical diagnostics, is the intrinsic fabrication-related polydispersity in particle sizes and magnetic properties. Therefore, to reduce this variation, it is prudent to integrate a pre-separation procedure for the particles into the overall workflow of the system. In this work, a concept for the controlled on-chip fractionation of micron-sized superparamagnetic beads (SPBs) is introduced, which is applicable for sorting magnetic particles according to their properties in a continuous operation mode. A specifically designed magnetic domain pattern is imprinted into an exchange-biased thin film system to generate a tailored magnetic stray field landscape (MFL), enabling lateral transport of SPBs when superposing the MFL with external magnetic field pulses. The domain pattern consists of parallel stripes with gradually increasing and decreasing width, resulting in a step-wise jumping motion of SPBs with increasing/decreasing jump distance. SPBs with different magnetophoretic mobilities, determined, among others, by the particle size and magnetic susceptibility, discontinue their lateral motion at different jump distances, i.e., different lateral positions on the substrate. Thorough analysis of the motion using optical microscopy and particle tracking revealed that an increasing stripe width not only leads to a larger jump distance but also to a lowered jump velocity. As a consequence, particles are spatially separated according to their magnetic and structural properties with a large throughput and time efficiency, as simultaneous sorting occurs for all particles present on the substrate using a constant sequence of short external field pulses.
\end{abstract}


\section{Introduction}\label{sec:intro}
Lab-on-a-chip (LOC) platforms or Miniaturized Total Chemical Analysis Systems ($\mu$-TAS) are considered key technologies for the implementation of rapid and cost-effective point-of-care sensing and diagnostics.\cite{Kricka2001,Knight2002,Manz1990} Existing solutions, like commonly used membrane-based lateral flow assays, are indeed cheaply produced devices, however, they come with drawbacks regarding sensitivity and reliability.\cite{Yetisen2013,Cardoso2017,Zhang2020} As a possibility for improvement, magnetic particles (MPs) in the size range of nanometers to micrometers are proposed as analytic probes.\cite{Pankhurst2003,Gijs2004,Pamme2006,Ruffert2016} Their surface properties are tunable via various chemical synthesis routes, offering substantial flexibility in terms of recognizing different analyte substances\cite{Gao2013,Moerland2019,Ran2014,Rampini2021,Reginka2021,Feely2023} and their magnetic characteristics can be exploited for a controlled actuation inside applied magnetic fields.\\ If a batch of spherical MPs shall be used on-chip, which is the typical application scenario, it is desirable that their magnetic field responses, leading to a certain motion velocity, do not differ too much, i.e., fall in a narrow range around an average value. The decisive quantity for magnetic field-controlled motion of spherical MPs in liquids is their magnetophoretic mobility $U_{\mathrm{m}}$, defined as
\begin{equation}
  U_{\mathrm{m}} = \frac{\Delta\chi V}{6\pi\eta R}
\label{eq:1}
\end{equation}
, where $\Delta\chi$ is the difference between the magnetic susceptibilities of MP and the surrounding liquid, $V$ is the volume of the MP, $\eta$ is the viscosity of the liquid and $R$ is the radius of the MP.\cite{Wise2015,Zhou2016} $U_{\mathrm{m}}$, therefore, depends on the effective MP size, the MP magnetic response to an applied magnetic field and the properties of the surrounding liquid. For a given liquid, the first two parameters may vary independently among the different MPs. Typically, for all MP batches, either commercially available or lab fabricated, exact sizes and magnetic susceptibilities are challenging to control in the fabrication process and, thus, may differ considerably. For defined on-chip functionalities, a presorting of MPs with respect to $U_{\mathrm{m}}$ is, therefore, a necessity. One concept to achieve such presorting is to use stepwise or ratchet-like MP motion technologies, where static magnetic field landscapes (MFLs) emerging from topographic micromagnetic structures on a surface\cite{Donolato2012,Rampini2016,Block2023} or stable magnetic domain patterns in flat films \cite{Tierno2009,Holzinger2015a,Urbaniak2024} are superposed by a time-varying external magnetic field. The MFLs contain large magnetic field gradients over micron-sized distances, resulting in comparably strong magnetic forces exerted onto the MPs. By dynamically transforming the field gradients via the superposed external field, a remote-controlled, directed MP motion close to the chip surface in a quiescent liquid is obtained, allowing for the implementation of various LOC functionalities.\cite{Yellen2007,Yellen2009,Holzinger2015a,Rampini2021,Huhnstock2024,Kim2025,Abedini-Nassab2025}\\ 
When applying trapezoidal field pulses\cite{Holzinger2015a}, given that the pulse length extends long enough in time, the MPs start to accelerate, reach their steady-state velocity during the pulse plateau, and slow down when reaching a neighboring position of minimal potential energy. The intra-pulse (maximum) steady-state velocity $v_{\mathrm{st}}$ is determined by the equilibrium of accelerating magnetic force and viscous drag exerted by the surrounding fluid.\cite{Liu2007,Holzinger2015a} For sufficiently short pulses, the change of the effective field (the vector sum of the static and the external field) acting on the MP is considerably faster than the time required by the MP to reach its next position with the given $v_{\mathrm{st}}$, effectively disabling a motion step for the MP. Hence, the pulse length (or pulse frequency) is a decisive experimental parameter for deciding whether MPs can follow the dynamic transformation of the effective magnetic field and, therefore, allow continuous MP motion. A higher frequency of pulses (or shorter pulse length) will only allow for motion of MPs with sufficiently large $U_{\mathrm{m}}$, resulting in a fractionation of the MP batch according to their magnetophoretic mobilities. As this fractionation concept necessitates sweeping the external field frequency, an autonomous, continuous operation within an LOC device is not straightforward to implement.\\
\begin{figure*}[ht!]
 \centering
 \includegraphics[width=12cm]{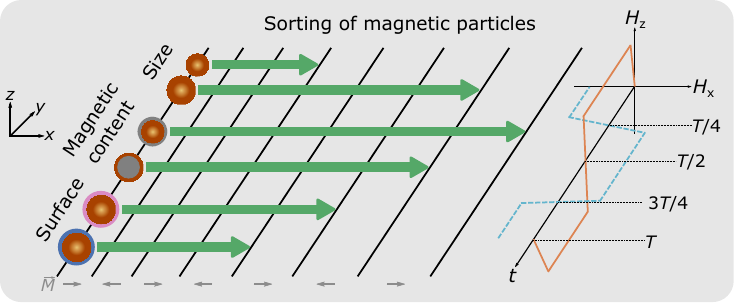}
 \caption{Schematic concept for the spatial fractionation of MPs in a microfluidic environment based on their directed transport above a magnetic stripe domain pattern with gradually increasing stripe width. MPs of varying characteristics (size, magnetic content distribution, surface properties) are expected to be immobilized at a locally different position along the domain pattern, depending on the respective stripe width. In this work, we experimentally demonstrate the feasibility of the fractionation concept by separating differently-sized MPs. The domains are magnetized as indicated by the grey arrows, resulting in a head-to-head (hh)/tail-to-tail (tt) magnetization configuration. A periodic sequence of trapezoidal magnetic field pulses in the vertical $z$-direction (solid, orange line) and the lateral $x$-direction (blue, dashed line) with a fixed alteration frequency is applied to achieve MP separation.}
 \label{fgr:1}
\end{figure*}
In this work, we present a unique MFL design based on an engineered magnetic domain pattern that enables localized MP fractionation using external field pulses of constant length. In particular, we will show how ion bombardment induced magnetic patterning (IBMP) of in-plane magnetized, exchange-biased layers can be employed to fabricate parallel stripe domains with variable widths. This domain pattern allows for the controlled spatial separation of superparamagnetic beads (SPBs) with differing $U_{\mathrm{m}}$ within a continuous experiment, where none of the external field parameters need to be adjusted (see Fig.~\ref{fgr:1}). This concept, therefore, possesses the potential to enable a straightforward time-efficient operation, which goes beyond the frequency-based separation method established in literature, e.g. \cite{Yellen2007,Rampini2021}, that typically focuses on highly symmetrical micromagnetic structures.

\section{Experimental}
\label{sec:experimental}
\subsection{Fabrication of magnetically patterned substrate}\label{subsec:substrate_fabrication}
A magnetic parallel-stripe domain pattern with gradually changing stripe width (stripe length equal to substrate size of ca.~\SI{10}{\milli\meter}) and a periodic head-to-head (hh)/tail-to-tail (tt) magnetization configuration was imprinted into an exchange-biased (EB) thin film system via IBMP.\cite{Mougin2001,Ehresmann2004,Ehresmann2015} The thin film system consisted of a Cu(\SI{5}{\nano\meter})/Ir$_{17}$Mn$_{83}$(\SI{30}{\nano\meter})/Co$_{70}$Fe$_{30}$(\SI{10}{\nano\meter})/Si(\SI{20}{\nano\meter}) layer stack, which was deposited onto a naturally oxidized Si(100) wafer piece (ca. \SI{10}{\milli\meter} $\times$ \SI{10}{\milli\meter}) by rf-sputtering at room temperature. Subsequently, the sample was field cooled to set the in-plane direction of the EB. For this, the sample was placed in a vacuum chamber (base pressure = 5~$\times$~10$^{-7}$~mbar) and annealed at \SI{300}{\degreeCelsius} for \SI{60}{\minute} in an in-plane magnetic field of \SI{145}{\milli\tesla}. In this state, the sample was ready for IBMP. Therefore, a homogeneous photoresist, with a sufficient thickness to prevent 10 keV He ions from penetrating the magnetic layers, was deposited on top of the thin film system via spin coating. The photoresist was structured to exhibit covered and uncovered areas of varying width and an equal length that corresponds to the size of the sample, using direct laser writing lithography. Starting with an uncovered stripe width of \SI{1.2}{\micro\meter}, the adjacent covered stripe had a width of \SI{1.5}{\micro\meter}, followed by another uncovered stripe of \SI{2}{\micro\meter}. This alternation of covered and uncovered areas with gradually increasing widths was repeated with an increment of \SI{0.5}{\micro\meter} until a stripe width of \SI{5}{\micro\meter} was reached. For this width, the alternation of covered and uncovered areas was repeated eleven times. In the following, covered and uncovered areas were further widened with an increment of \SI{0.5}{\micro\meter} until a width of \SI{10}{\micro\meter} was finally reached. These widest covered and uncovered resist areas were repeated seven times. Finally, the width of covered/uncovered areas was decreased back to the starting value of \SI{1.2}{\micro\meter}, following the same incrementation parameters as for the width increase. This increase/decrease procedure is repeated periodically throughout the whole substrate area. Fig.~S1 in the Supplementary Information presents a local microscope image of the resist structure design. The long axis of the stripes was oriented perpendicular to the initial EB direction (set by the field cooling procedure). After resist structure fabrication, the sample was bombarded with a dose of \SI{1e15}{\per\centi\meter\squared} He ions (kinetic energy of \SI{10}{\kilo\electronvolt}) employing a home-built Penning ion source.\cite{Lengemann2012} For a periodic hh/tt magnetization of adjacent stripe domains, an in-plane homogeneous magnetic field (\SI{100}{\milli\tesla}) was applied during ion bombardment, pointing antiparallel to the direction of the EB initializing field employed in the field cooling procedure. After bombardment, the photoresist was removed by washing the sample thoroughly with acetone. Finally, the sample surface was cleaned by rinsing it with acetone, isopropanol, and water. After drying, a \SI{500}{\nano\meter} thick Poly(methyl methacrylate) (PMMA) spacing layer was deposited on top of the sample via spin coating.
\subsection{Transport of superparamagnetic beads}\label{subsec:bead_transport}
For inducing the directed motion of SPBs, 20 µL of a diluted aqueous dispersion of SPBs (Dynabeads M-270 Carboxylic Acid/Dynabeads MyOne) was pipetted on top of the magnetically patterned substrate. The fluid containing the SPBs was confined by a microfluidic chamber that was adhered to the substrate surface. The chamber was produced by cutting a square of approximately $\SI{8}{\milli\meter}\times\SI{8}{\milli\meter}$ out of a $\SI{10}{\milli\meter}\times\SI{10}{\milli\meter}$ sized Parafilm sheet. The chamber was sealed by a square-shaped glass coverslip and the substrate with the SPBs on top was placed in the center of a Helmholtz coil arrangement. Within this arrangement, orthogonally placed Helmholtz coils allowed for the application of homogeneous trapezoidal magnetic field pulses perpendicular and parallel to the transport substrate plane, i.e., the $z$- and $x$-direction. The substrate itself was aligned with the substrate plane normal pointing parallel to the $z$-direction and the in-plane magnetization direction pointing parallel to the $x$-direction (see Fig.~\ref{fgr:1}). Each magnetic field pulse consisted of a linear rising time, a plateau time, and a linear drop time. The rising and drop times were determined by the pulse magnitude $H_{\mathrm{max}}$ as well as the alteration rate of the external magnetic field being \SI{3.2e6}{\ampere\per\meter\per\second}. The pulse magnitudes were chosen to be $\mu_{0}H_{\mathrm{max,x}}=\SI{1}{\milli\tesla}$ for the $x$-direction field and $\mu_{0}H_{\mathrm{max,z}}=\SI{1}{\milli\tesla}$ or \SI{2}{\milli\tesla} for the $z$-direction field. For the initialization of directed SPB motion, a temporal phase shift of $\pi/2$ between pulse sequences in the $z$- and $x$-direction and a periodic change of pulse orientation between $H_{\mathrm{max}}$ and $-H_{\mathrm{max}}$ was implemented (see Fig.~\ref{fgr:1}). For observing and recording the SPB motion, the sample was approached with an optical bright-field microscope in reflection mode. SPB motion was recorded with an attached high speed camera: Both a Mikrotron EoSens CoaXPress CXP-6 camera (maximum resolution of 4096 px $\times$ 3072 px) at a framerate of 25 frames per second (fps) and an Optronis CR450×2 camera (maximum resolution of 800 px $\times$ 600 px) at a framerate of 1000 fps were employed for SPB transport characterization.
\section{Results}\label{sec:results}
\paragraph{Qualitative investigation}
\begin{figure*}[ht!]
 \centering
 \includegraphics[width=\textwidth]{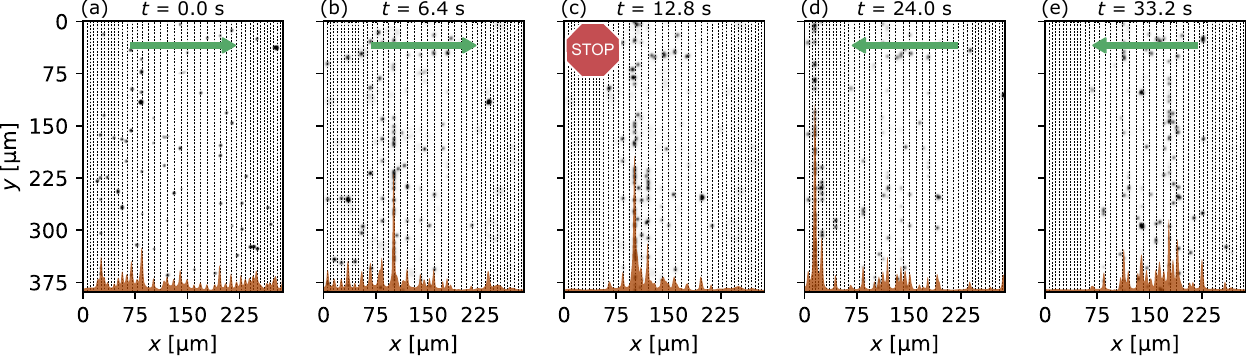}
 \caption{Optical analysis for the motion behavior of SPBs dispersed in water on top of a magnetic stripe domain pattern with gradually varied stripe width. Shown are single optical microscope images (a)-(e) obtained at different recording times $t$ of a video recording taken with a frame rate of 25 fps. For the recorded experiment, a periodic sequence of trapezoidal magnetic field pulses with $T/4$~=~\SI{0.1}{\second} and $\mu_{0}H_{\mathrm{max,x}}~=~\mu_{0}H_{\mathrm{max,z}}~=~\SI{1}{\milli\tesla}$ was applied. The stripe domain pattern within the underlying substrate is indicated in each displayed frame by black dashed lines. For frames (a) and (b), SPBs (black spots) are transported from left to right across the substrate (see green arrows), leading eventually to immobilization at a lateral $x$-position of ca. \SI{100}{\micro\metre}, prominently visible in (c). Here, the external field sequence was stopped and paused for \SI{8}{\second}, resulting in a halted SPB motion. Upon re-initialization of the external field sequence with inverted phase relation between field pulses in $z$- and $x$-direction, SPBs are transported into the opposite direction, emphasized in (d) and (e) by the green arrows. Frame (e) highlights the immobilization of SPBs this time around ca. \SI{200}{\micro\metre} $x$-position. Lateral particle density profiles, averaged along the $y$-dimension, are included as brown-filled curves in each frame.}
 \label{fgr:2}
\end{figure*}
SPBs with a diameter of \SI{2.8}{\micro\meter} were placed on top of an EB thin film substrate magnetically patterned by IBMP. The pattern consists of parallel stripe domains with gradually increasing/decreasing stripe width and a periodic in-plane head-to-head (hh) and tail-to-tail (tt) magnetization configuration (see Fig.~\ref{fgr:1}). An aqueous dispersion of the SPBs was pipetted onto the magnetically patterned substrate (covered by a \SI{500}{\nano\meter} thick PMMA layer) and a sequence of trapezoidal magnetic field pulses in $z$- and $x$-direction (see Fig.~\ref{fgr:1}) was applied after letting the SPBs sediment towards the substrate surface. Fig.~\ref{fgr:2}(a) presents a microscope image taken before applying the external fields ($t~=~\SI{0.0}{\second}$). For better recognition of SPBs, the background of the image, containing inhomogeneous illumination and substrate defects, was subtracted and a denoising algorithm was applied. SPBs are statistically distributed across the field of view in this initial state. The approximated positions of stripe domain walls (DWs) in the underlying substrate are indicated by dashed black lines. For the lateral $x$-coordinate, SPBs are mostly located directly above or very close to a DW, owing to the attraction toward maximum stray field strength. Upon initiating an external magnetic field pulse sequence with a period of $T~=~\SI{0.4}{\second}$ and pulse amplitudes of $\mu_{0}H_{\mathrm{z,ext}}~=~\mu_{0}H_{\mathrm{x,ext}}~=~\SI{1}{\milli\tesla}$ in $z$- and $x$-direction, respectively, SPBs performed a stepwise motion in positive $x$-direction, towards increasing stripe domain width, i.e., increasing DW separation distances. Reaching a certain DW separation distance, SPBs began to perform oscillating movements along the lateral $x$-dimension, effectively losing their transportability and being, therefore, classified as immobile. SPBs located at the substrate area of maximum DW separation directly after sedimentation showed this oscillatory motion from the start of the external field sequence. At $t~=~\SI{6.4}{\second}$, the gradual SPB immobilization was observable by an increasing accumulation of SPBs at a lateral position of $x\approx\SI{100}{\micro\meter}$ (Fig.~\ref{fgr:2}(b)) though a few SPBs were immobilized even earlier, starting from around \SI{60}{\micro\meter} $x$-position. Here, SPBs are prominently caught between \SI{9.0}{\micro\meter} and \SI{9.5}{\micro\meter} wide stripe domains. After $t~=~\SI{12.8}{\second}$, all SPBs moving from left to right are immobilized, leading to a maximum accumulation of SPBs at the largest DW separations (Fig.~\ref{fgr:2}(c)). The external fields were stopped at this moment. Probing whether SPBs can regain their transportability, a pulse sequence with inverted phase relation between fields in $z$- and $x$-direction was applied after a pause, initiating SPB transport in the opposite direction. As a consequence, formerly immobilized SPBs at $\approx\SI{100}{\micro\meter}$ $x$-position could be transported from right to left, as can be seen for $t~=~\SI{24.0}{\second}$ in Fig.~\ref{fgr:2}(d). At the end of the experiment, at $t~=~\SI{33.2}{\second}$, SPBs were accumulated at an $x$-position of $\approx\SI{200}{\micro\meter}$. They represent SPBs that were previously immobilized at larger DW separation distances not seen in the field of view, owing to the periodic repetition of increasing/decreasing stripe domain width. After inverting the phase relation for the external field sequence, they were transported back into the field of view and finally immobilized again at larger DW separation distances.\\
\begin{figure*}[t!]
 \centering
 \includegraphics[width=\linewidth]{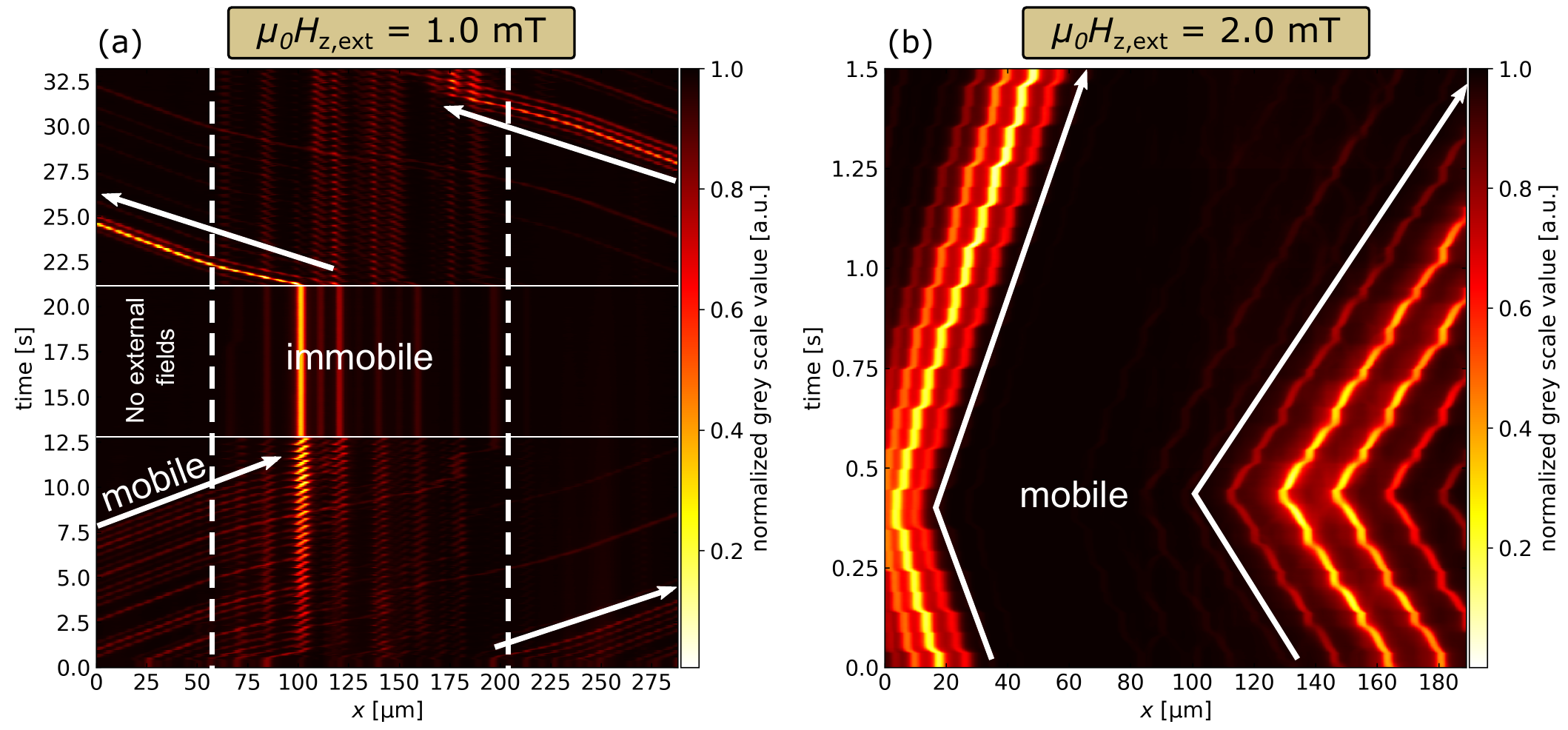}
 \caption{Motion behavior of SPBs dispersed in water on top of a magnetically patterned substrate with increasing stripe domain width for two different external magnetic field amplitudes in $z$-direction. A qualitative impression of the SPB transport is given by presenting averaged lateral intensity profiles of each recorded image frame as a function of recording time. The color scale indicates the image intensity; high intensity represents the background and low intensity the presence of SPB rows. For weak external magnetic field pulses of $\mu_{0}H_{\mathrm{z,ext}}~=~\mu_{0}H_{\mathrm{x,ext}}~=~\SI{1.0}{\milli\tesla}$ magnitude (a), fractions of mobile and immobile SPBs were distinguishable in dependence on the substrate position. The immobilization of SPBs occurred at a substrate area between $\approx\SI{60}{\micro\meter}$ and $\approx\SI{200}{\micro\meter}$ lateral $x$-position (marked by dashed white lines), coinciding with the substrate region of largest stripe domain width. Mobile SPBs were initially transported from left to right until immobilization. After a pause, the external field sequence was modified so that SPB transport from right to left was initiated. Most of the previously mobile SPBs regained their mobility by this. Stronger field pulses of $\mu_{0}H_{\mathrm{z,ext}}~=~\SI{2}{\milli\tesla}$ magnitude (b) lead to mobilization of SPBs even for the region of largest stripe domain width.}
 \label{fgr:3}
\end{figure*}
From the initial experiment, first evidence for the fractionation capability of the investigated domain pattern design is collected. For studying the SPB motion behavior in more detail, continuous tracking of SPB rows was realized by looking at the time dependence of lateral intensity profiles, as a measure for SPB density, within each recorded image frame of the transport experiment. These profiles represent the image pixel grey scale values along the $x$-dimension and were averaged for each image along the $y$-direction to get the mean position information for each SPB row. Examples of the intensity profiles are documented as brown-filled curves in each image of Fig.~\ref{fgr:2}(a) - (e). Peaks within the profiles mark the momentary lateral positions of single SPB rows. Plotting the lateral intensity profiles of all images that were recorded during the experiment as a function of time creates a 2D false-color image that enables the tracing of each SPB row position throughout the experiment. Using this evaluation method, the qualitative SPB motion behavior can be visualized more intuitively and the influence of changing experimental parameters can be detected at first glance. For instance, Fig.~\ref{fgr:3}(a) shows the map of lateral intensity profiles in dependence on experimental time for the initially described transport experiment of Fig.~\ref{fgr:2}. Continuous lines of high intensity (low grey scale value) mark the trajectories of single SPB rows, whereby the absolute grey scale value correlates antiproportionally with the number of particles present in a specific row.\\
Initially, SPBs are transported from left to right through the field of view, leading to diagonal intensity lines, except for the substrate region, where SPBs perform oscillating movements from the beginning. This is the case for a region between $\approx\SI{60}{\micro\meter}$ and $\approx\SI{200}{\micro\meter}$ lateral $x$-position, which is therefore highlighted in Fig.~\ref{fgr:3}(a) by two dashed white lines. Over time, SPBs are either approaching this region from the left or leaving this region to the right. For most of the approaching SPBs, it could be observed that they eventually transition to oscillating movement, most prominently before the stripe domain width increases to \SI{9.5}{\micro\meter}. This is mirrored in the intensity map of Fig.~\ref{fgr:3}(a) by a zig-zagging line at $\approx\SI{100}{\micro\meter}$ $x$-position, with an increase of intensity (decrease of grey scale value) observable for this line over time. Thus, an increasing number of approaching SPBs was immobilized at this position, accounting for the rising SPB density. Some SPBs transition to oscillating motion at larger stripe domain widths or even continue their directed motion, i.e., they are transported beyond the immobilization region. These SPBs acquire larger motion velocities, most likely due to a larger magnetophoretic mobility (see Eq.~\ref{eq:1}). Hence, they are enabled even for the largest stripe domain width (\SI{10}{\micro\meter}) to traverse the distance between two neighboring DWs in phase with the transformation of the MFL by the external fields. This physical mechanism will be analyzed in more detail in the Discussion section. Stopping the external field sequence at $t\approx\SI{13}{\second}$ (marked by the lower horizontal white line in Fig.~\ref{fgr:3}(a)), the SPBs rest at their immobilized position, indicated by straight lines in the intensity map. After $t\approx\SI{21}{\second}$ (marked by the upper horizontal white line in Fig.~\ref{fgr:3}(a)), an adjusted field sequence leads to transport of immobilized SPBs towards the left edge of the field of view. It is most prominently visible that almost all SPBs immobilized at $x\approx\SI{100}{\micro\meter}$ regained their transportability and could be relocated. SPBs that stay in the immobilized region are mostly those that were already located there from the start of the experiment, i.e., after sedimentation. They supposedly don't possess sufficient magnetophoretic mobility to traverse the respective DW separation distance. Mobile SPBs that approached the region of largest stripe domain width from the right yielded similar observations: They either transitioned to oscillatory motion at a certain DW separation distance or, in a few rare cases, maintained their directed motion. A large group of SPBs entering the field of view at $t\approx\SI{27.5}{\second}$ was prominently immobilized at $x\approx\SI{175}{\micro\meter}$. These SPB rows arrived in close succession since they were immobilized according to their respective magnetophoretic mobility at large stripe domain widths outside of the chosen field of view.\\ 
For studying the influence of the external field strength on the SPBs' transportability, a similar experiment was conducted with a magnitude of $\mu_{0}H_{\mathrm{z,ext}}~=~\SI{2}{\milli\tesla}$ for the field pulses in $z$-direction, while the pulses in $x$-direction were kept at $\mu_{0}H_{\mathrm{x,ext}}~=~\SI{1}{\milli\tesla}$ magnitude. The result for a \SI{1.5}{\second} long segment of this experiment is displayed in Fig.~\ref{fgr:3}(b). Two fractions of SPBs are visible, one at the right edge and one on the left edge of the field of view. The right fraction is located at larger DW separation distances than the left fraction. Remarkably, even the fraction at the largest DW separation distances shows continuous transportability in both directions, which was not true for the lower external field strength. For this particular experiment, even a decreased period $T~=~\SI{0.2}{\second}$ still allowed for continuous SPB transport throughout the whole substrate area, owing to the increased external field strength. The transportability of the right fraction must, therefore, be connected to a larger magnetic force acting on the SPBs with the increased magnitude of $H_{\mathrm{z,ext}}$. With a larger driving force, SPBs gain in their motion velocity and are consequently enabled to follow the transformation of the MFL in phase. To quantify this increase in motion velocity, single SPBs were measured to achieve an average maximum velocity (acquired during a transport step) of $\SI{150 \pm 20}{\micro\meter\per\second}$ for \SI{1}{\milli\tesla} $H_{z}$ pulse magnitude while for \SI{2}{\milli\tesla} the velocity sprung up to $\SI{220 \pm 20}{\micro\meter\per\second}$. The results suggest that the external field strength is a regulator for controlling SPB mobility towards specific applications.\\
\begin{figure*}[ht!]
 \centering
 \includegraphics[width=\linewidth]{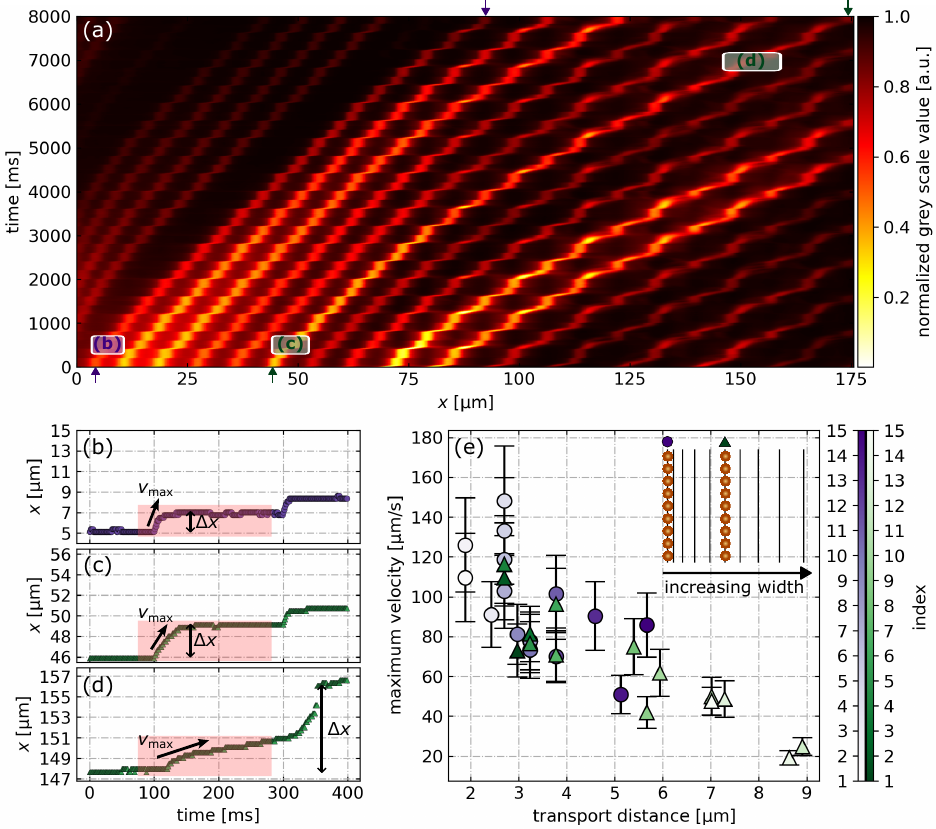}
 \caption{Quantitative analysis of SPB motion dynamics in water above a magnetically patterned substrate with increasing stripe domain width. Starting from the evolution of averaged lateral intensity profiles acquired from each recorded image frame during the experiment (a), three SPB row trajectory snippets highlighted with white-framed rectangles were picked as examples for the identification of maximum particle velocity in (b) - (d). The period of the external magnetic field pulse sequence was chosen to be $T~=~\SI{0.8}{\second}$ with pulse magnitudes of $\mu_{0}H_{\mathrm{z,ext}}~=~\mu_{0}H_{\mathrm{x,ext}}~=~\SI{1}{\milli\tesla}$ to initiate SPB transport from left to right throughout all domain widths present in the underlying substrate (no immobilization). Maximum SPB row motion velocities, as a measure for their averaged mobility, were determined for the transport step initiated after changing the sign of the external field in $z$-direction. The red-marked areas for the exemplary trajectory data $x(t)$ in (b) - (d) were considered for the velocity calculation by fitting a Gaussian error function to the marked data and subsequently obtaining the velocity as the maximum of the fit's time derivative. Each velocity is assigned a transport step distance $\Delta x$, which corresponds to the difference between two consecutive plateaus in the trajectory data (see further explanation in the text). Plotting the obtained velocities as a function of the respective transport distance for the two SPB row trajectories that contain the exemplary steps of (b) - (d), a decreasing tendency for the velocity with larger transport steps (larger domain width) is observable (e).}
 \label{fgr:4}
\end{figure*}
\paragraph{Quantitative analysis of step velocity}
A deeper analysis uncovered an additional dependency of the SPB steady-state velocity (the velocity acquired by an SPB after its acceleration phase) on the current lateral position above the substrate. More specifically, it depends on the distance traveled by the SPB during one motion cycle, i.e., the DW separation distance. The analysis was carried out by initiating SPB movement over the whole substrate area, thus inducing transport steps for all DW separation distances. The following parameters for the external magnetic field pulse sequence were, therefore, chosen in order to avoid immobilization of the SPBs: $\mu_{0}H_{\mathrm{z,ext}}~=~\SI{1}{\milli\tesla}$, $\mu_{0}H_{\mathrm{x,ext}}~=~\SI{1}{\milli\tesla}$ and $T~=~\SI{0.8}{\second}$. The increase in $T$ gave the SPBs sufficient time to follow the transformation of the MFL in phase and thereby maintain continuous transportability. Microscope images of the experiment were recorded for \SI{8}{\second} at a frame rate of 1000~fps and averaged lateral particle density profiles were determined for every image. The evolution of these profiles over time is shown in Fig.~\ref{fgr:4}(a), indicating transport of SPB rows from left to right through the field of view. From a qualitative point of view, small transport steps within the SPB row trajectories can be observed for lower $x$-positions, while with larger $x$-position the distance per step is increasing. The trajectories of 19 SPB rows are traceable in total for this intensity profile map. Two trajectories were handpicked to serve as exemplary trajectories for the following description and visualization of steady-state velocity determination and its dependency on the SPB transport step distance. The starting and end positions of these trajectories are marked with purple and green arrows in Fig.~\ref{fgr:4}(a), respectively. They were chosen because a wide range of DW separation distances from \SI{2}{\micro\meter} to \SI{10}{\micro\meter}, i.e., a wide range of SPB transport step distances are covered herein. It needs to be emphasized at this point that no single particle analysis was taken into account, for the reason that single particle tracking was not feasible for the experimentally obtained SPB densities. As we were interested in the general development of SPB steady-state velocity with increasing stripe domain width, the tracking of SPB rows via lateral intensity profiles sufficed.\\ 
The SPB row motion consists of two distinguishable phases: A transport phase towards the adjacent DW that is induced with the application of magnetic field pulses in $z$ and $x$-direction, as well as a resting phase where the motion of the SPBs is completed and the application of the next field pulse is awaited. Both phases can be observed more closely by showing \SI{400}{\milli\second} long snippets of the exemplary trajectories in Fig.~\ref{fgr:4}(b) - (d). The position of each snippet within the overall SPB row trajectory is marked by white shaded areas in Fig.~\ref{fgr:4}(a). Note that the visualization is now reversed, plotting the $x$-position data for intensity peaks in the map of Fig.~\ref{fgr:4}(a) as a function of time $t$. Each snippet contains a larger transport step, which results from applying a field pulse in the $z$-direction, and a smaller transport step, which is the consequence of a field pulse in the $x$-direction (comparable to previous studies for hh/tt stripe domains of equal widths~\cite{Holzinger2015a,Reginka2021}). The resting phase of SPBs is signified by plateaus in the $x(t)$ data. The larger transport step was used to determine the steady-state velocity of SPBs (averaged for all SPBs residing in the same row) during one motion event and the accompanying transport step distance. For doing so, each larger step was placed into a $x(t)$ slice of \SI{200}{\milli\second} as indicated by the red shaded areas in Fig.~\ref{fgr:4}(b) - (d). A Gaussian error function was used as an approximation for the $x(t)$ data during the transport step and therefore fitted to each slice. A distribution of velocities over time $v(t)$ was retrieved by computing the derivative of the obtained fit function. The maximum $v_{\mathrm{max}}$ of the distribution was chosen as a measure for the steady-state velocity of the respective SPB row. The steady-state velocity is governed by an equilibrium of magnetic and friction force, theoretically leading to trajectory data of constant slope. As there are no regions of constant slope within the Gaussian error fit function, the obtained maximum velocity can only be an approximation for the real steady-state velocity. This is taken into consideration for the error of the velocity, which was determined by averaging over all values present above the full width at half maximum (FWHM) of the obtained distribution and calculating the difference of this averaged velocity to $v_{\mathrm{max}}$ (see Discussion section).\\
Qualitative changes in $v_{\mathrm{max}}$ can be made out when comparing the trajectory snippets of Fig.~\ref{fgr:4}(b) - (d), where each snippet represents a different transport step distance. The transport step distance was characterized as the difference $\Delta x$ between two adjacent $x(t)$ plateaus. After initiation of the larger transport step at $t~=~\SI{100}{\milli\second}$, the next plateau is reached after a few milliseconds for the exemplary data shown in Fig.~\ref{fgr:4}(b) and (c), indicating a comparably fast SPB row motion. The transport step distance $\Delta x$ is increasing from (b) to (c), the steady-state velocity $v_{\mathrm{max}}$, however, appears to decrease. This trend continues for the exemplary data of Fig.~\ref{fgr:4}(d) that represents SPB motion above the substrate area with maximum DW separation distance. Remarkably, the SPB row motion is induced with a delay when compared to the previous transport steps, starting around \SI{110}{\milli\second} instead of \SI{100}{\milli\second} (a physical explanation is provided in the Discussion section). The ensuing motion is visibly slower and also does not come to a full stop before $t~=~\SI{300}{\milli\second}$, where an external field pulse in the $x$-direction is applied. This induces the smaller transport step, which can be identified more clearly in Fig.~\ref{fgr:4}(b) and (c). In these instances, the SPB rows have already come to rest above the adjacent DW position before the initiation of the smaller step. For the step presented in Fig.~\ref{fgr:4}(d), the SPB row has not yet reached the position of the adjacent DW. Determining $\Delta x$ at this time would, therefore, not reflect the corresponding separation distance of neighboring DWs in the underlying substrate for this particular transport step. The additional transformation of the MFL by an applied pulse in the $x$-direction triggers another acceleration phase for the SPB row. The SPBs finally come to rest after around $t~=~\SI{350}{\milli\second}$, marking the approximate location of the adjacent DW and therefore making it possible to quantify $\Delta x$ for larger DW separation distances. For the sake of comparability, $v_{\mathrm{max}}$ is determined in this case only for the first induced motion phase, as indicated in Fig.~\ref{fgr:4}(d). Evaluating $v_{\mathrm{max}}$ for every larger transport step within the two selected SPB row trajectories in dependence on the transport step distance $\Delta x$ yields the result shown in Fig.~\ref{fgr:4}(e).\\
Data points measured for the first SPB row (starting at the substrate area of lowest DW separation distance) are represented by purple circles and data points measured for the second SPB row (starting at the region of larger DW separation distances) by green triangles. The color scale for the data points indicates the transport step index, i.e., the occurrence of the transport step within the analyzed trajectory. The lower the index, the earlier the transport step was observed (typically at a lower DW separation distance). The correlation between $v_{\mathrm{max}}$ and $\Delta x$ reflects the qualitative trend: With increasing transport distance of the SPBs (DW separation distance) the steady-state velocity is lowered, from over \SI{140}{\micro\meter\per\second} for the lowest transport distances down to \SI{20}{\micro\meter\per\second} for the largest steps. It has to be noted that not all transport steps within the two trajectories could be evaluated for $v_{\mathrm{max}}$ due to insufficient fit quality. Therefore, some data points are missing, most prominently for the largest transport distances, which, however, does not lessen the clarity of the observed trend. It becomes even more evident when looking at the steady-state velocities obtained from the analysis of all visible SPB row trajectories shown in Fig.~S2 in the Supplementary Information. In terms of SPB fractionation efficiency, the beads' mobility is significantly decreased for larger domain widths in the underlying substrate due to the lowered motion velocity, leading to the observed oscillatory behavior for smaller periods of the external field sequence. A physical explanation for this correlation will be approached in the Discussion section.
\paragraph{Fractionation of differently-sized beads}
\begin{figure*}[t!]
 \centering
 \includegraphics[width=15cm]{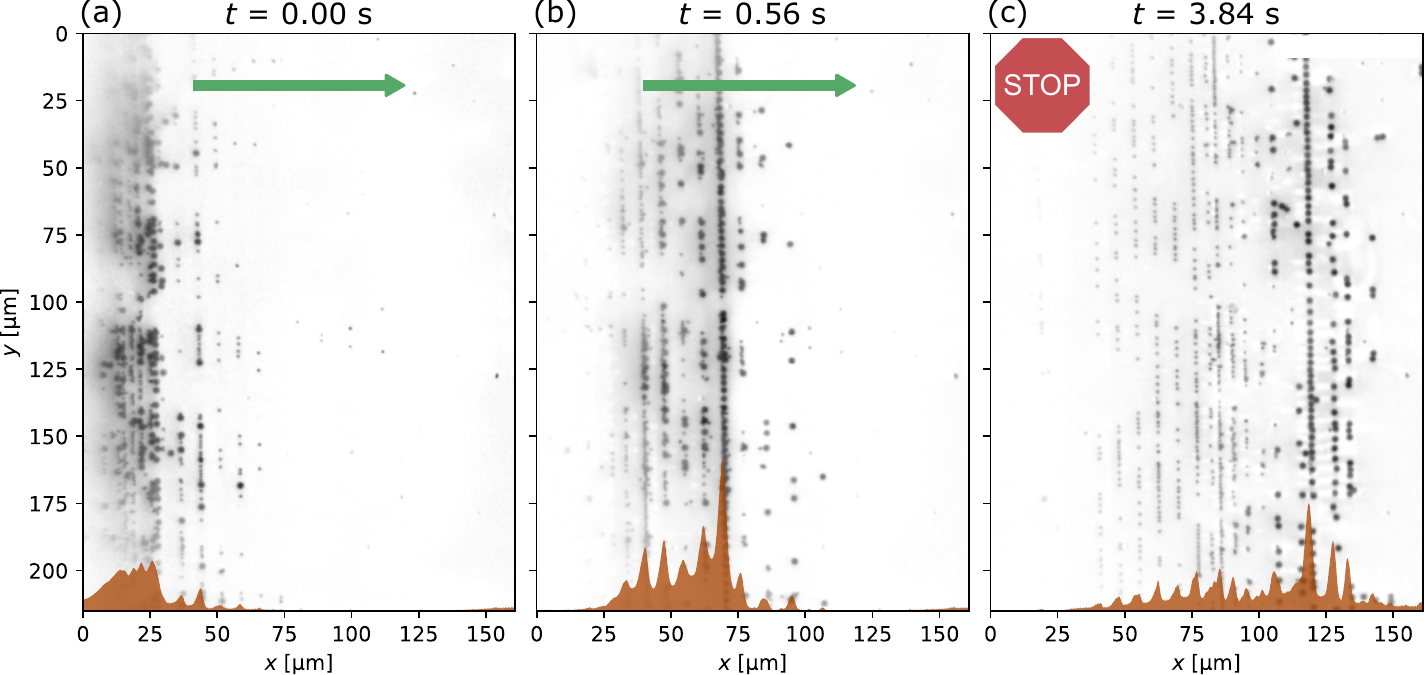}
 \caption{Fractionation of differently sized SPB populations on top of a magnetically stripe domain patterned substrate with increasing stripe width. (a) All SPBs are assembled on the left edge of the microscope's field of view, resembling minimal stripe domain width in the underlying substrate. (b) Upon application of an external magnetic field pulse sequence of constant frequency, all SPBs are transported to the right side of the field of view. (c) After a short time, the SPBs are immobilized (oscillatory motion) at specific locations. Statistically, smaller SPBs ($d~=~\SI{1}{\micro\meter}$) are immobilized earlier (at lower lateral $x$-position) than larger SPBs ($d~=~\SI{2.8}{\micro\meter}$). The two SPB populations were spatially separated into distinguishable fractions. Averaged intensity profiles (brown-filled curves) indicate the lateral extent of each fraction above the substrate surface.}
 \label{fgr:5}
\end{figure*}
According to Eq.~\ref{eq:1}, the magnetophoretic mobility of an MP transported by magnetic field gradients within a liquid environment is determined by its hydrodynamic radius and magnetic moment, given that the properties of the liquid stay constant. Therefore, we probed the fractionation of MP species with significantly different magnetophoretic mobilities by observing a mixture of differently-sized SPBs in the same experiment. Hence, an aqueous mixture of Dynabeads MyOne ($d~=~\SI{1}{\micro\meter}$) and Dynabeads M-270 ($d~=~\SI{2.8}{\micro\meter}$) was prepared and placed on top of the substrate with varying magnetic stripe domain width covered by a \SI{500}{\nano\meter} thick PMMA layer. To initiate the fractionation experiment, the majority of all present SPBs were transported to the substrate area of smallest stripe domain width, choosing the appropriate external magnetic field sequence. The achieved distribution of SPBs before the start of the experiment ($t~=~\SI{0}{\second}$) is shown in Fig.~\ref{fgr:5}(a). An averaged lateral intensity profile (brown-filled curve) is included to highlight the position of SPB rows and the number of SPBs per row. For this starting condition, SPBs of all sizes are assembled in rows on top of the DWs in the underlying substrate with an overall random distribution of $d~=~\SI{1}{\micro\meter}$- and $d~=~\SI{2.8}{\micro\meter}$-sized beads within each row. A sequence of external magnetic field pulses ($\mu_{0}H_{\mathrm{z,ext}}~=~\SI{2}{\milli\tesla}$, $\mu_{0}H_{\mathrm{x,ext}}~=~\SI{1}{\milli\tesla}$, $T~=~\SI{0.08}{\second}$) was applied, with the phase relation between pulses in $z$- and $x$-direction defined so that SPBs are moving from the left to the right through the field of view. Due to the relatively short period of the applied sequence, SPBs are traversing the imaged substrate area in under one second, emphasized by the snapshot taken at $t~=~\SI{0.56}{\second}$ in Fig.~\ref{fgr:5}(b). At this point, the differently sized SPBs are moving simultaneously and their spatial distribution is, therefore, still statistically defined. This changes after a few seconds: All SPBs are immobilized (oscillatory motion around a DW position) with an observable separation of the two SPB species at different substrate areas. The image shown in Fig.~\ref{fgr:5}(c) marks the end of the experiment at $t~=~\SI{3.84}{\second}$ and highlights the final spatial distribution of the SPBs. While \SI{1}{\micro\meter} sized beads are most prominently located at $x$-positions between \SI{40}{\micro\meter} and \SI{100}{\micro\meter}, \SI{2.8}{\micro\meter} sized beads reached further across the substrate and were immobilized at $x$-positions between \SI{100}{\micro\meter} and \SI{150}{\micro\meter}. Overall, Dynabeads M270 proved to have higher magnetophoretic mobility, as known from literature \cite{Grob2018}, therefore being enabled to traverse larger DW separation distances than Dynabeads MyOne. Analogous to the previous observations for nominal equally sized SPBs, all beads of one species are not immobilized above one specific DW but are rather spread within an immobilizing region. Judging from the peak heights within the averaged lateral intensity profile shown in Fig.~\ref{fgr:5}(c), which correlate with the number of SPBs present within one specific row, the majority of Dynabeads MyOne was immobilized at \SI{90}{\micro\meter} $x$-position and the majority of Dynabeads M-270 at \SI{120}{\micro\meter}. Supposedly, the reason for the observed broad distribution of SPBs lies within a statistical fluctuation of bead size around the nominal value. Typically, commercially available batches of SPBs show some degree of polydispersity. In addition, some areas of the substrate consist of a repetition of \SI{5}{\micro\meter} and \SI{10}{\micro\meter} wide stripe domains. Here, minor variations in the local effective magnetic field and/or substrate surface properties might lead to SPB immobilization above different DWs. On average, however, the conducted proof-of-principle experiment emphasizes the capability of our MFL (domain pattern) design to create spatially separated fractions of MP species with significantly differing magnetophoretic mobilities. The fractions can be collected for further analysis by transporting them simultaneously, while still being separated, via magnetic field pulses with increased $H_{z}$ magnitude (see results of Fig.~\ref{fgr:3}(b)).
\section{Discussion}\label{sec:discussion}
The observed immobilization of SPBs for larger DW separation distances in the underlying magnetically patterned substrate can be traced back to two physical mechanisms. On the one hand, SPBs must traverse larger distances when a transport step is induced, effectively increasing the time $t_{\mathrm{step}}$ required for reaching the adjacent DW position when assuming a constant average velocity. If $t_{\mathrm{step}}$ is larger than a certain critical length of time $t_{\mathrm{crit}}$, SPBs are more likely to slip back to their original position before the initialization of the next transport step, therefore, entering the so-called "phase-slipping" transport mode \cite{Yellen2007}. Both the literature and our study highlight that the period $T$ of the external field sequence has a large impact on $t_{\mathrm{crit}}$. In this work, we demonstrated that $t_{\mathrm{step}}$ can be deliberately adjusted by the separation distance of artificial DWs imprinted into an in-plane EB thin film system. As a result, the transition from directed SPB transport ("phase-locked" regime) to oscillating motion ("phase-slipping" regime) is a function of SPB position at constant $T$, i.e., constant $t_{\mathrm{crit}}$. On the other hand, the motion velocity of same-sized SPBs is not independent of the DW separation distance, as the quantitative investigation shown in Fig.~\ref{fgr:4} and Fig.~S2 in the Supplementary Information has demonstrated. Larger DW separation distances lead to lower maximum SPB velocities after the application of an external $z$-field pulse. This further contributes to an increased $t_{\mathrm{step}}$ and a higher probability for an SPB to switch to the "phase-slipping" regime. The observed trend itself is very likely to originate either from a decreasing accelerating magnetic force or an increasing decelerating friction force. For assessing the magnetic force, the static MFL on top of the substrate was simulated in a two-step process: First, the micromagnetic simulation package MuMax3\cite{Vansteenkiste2014} was employed to compute the magnetization distribution within the domain pattern design. In the second step, emerging stray fields were calculated in a dipole approximation according to the obtained magnetization distribution. The results for the stray field components $H_{\mathrm{z}}$ and $H_{\mathrm{x}}$ at a distance $z~=~\SI{2.5}{\micro\meter}$ above the magnetic thin film system together with further details on the simulations are given in the Supplementary Information (see Fig.~S3 and S4). Starting at a domain width of \SI{1}{\micro\meter} and increasing the width gradually until \SI{5}{\micro\meter}, the results suggest an increasing stray field strength on top of the respective DWs. As an SPB's magnetic moment is proportional to the applied magnetic field strength, this would mean an increasing magnetic force with larger domain width, which seems to contradict the experimental observation. It does, however, provide a possible explanation for the delay in motion initialization observed for larger stripe domain widths (see Fig.~\ref{fgr:4}(d)). With increasing magnetic attraction towards the underlying polymer surface, SPBs might get close enough for attractive van-der-Waals interactions to become more influential. In this case, the observed delay could be caused by an initial adherence of the SPBs to the polymer surface, eventually overcome by the magnetic force pushing the particles to the next DW location. What has not been considered at this point is the magnetic field gradient, determining the acting magnetic force together with the SPBs' magnetic moments. Confining the stray fields in a smaller volume (smaller DW separation distance) may lead to an increased field gradient and, in turn, an increasing magnetic force. This tendency could, however, be balanced out by the lower field strength for the smallest domain widths.\\
Finally, we may consider different hydrodynamic drag situations in dependence on the respective DW separation distance. For the smallest stripe domain width, where the stray field strength is the lowest, SPBs are expected to exhibit an increased separation distance from the underlying substrate surface due to lessened magnetic attraction. The friction force for a spherical particle moving close to a flat surface inside a fluid is, among others, a function of a $z$-position-dependent friction coefficient \cite{Wirix-Speetjens2005}. The correlation is hereby anti-proportional: With increasing distance $z$ from the surface, the friction coefficient is decreased. The resulting reduced drag may partially explain the observed larger SPB motion velocities for the smallest transport step distances.\\
It was generally observed that for the chosen domain pattern design and a period $T~=~\SI{0.4}{\second}$ for the sequence of \SI{1}{\milli\tesla} strong magnetic field pulses, most of \SI{2.8}{\micro\meter}-sized SPBs were immobilized at a DW position between \SI{9}{\micro\meter} and \SI{9.5}{\micro\meter} wide stripe domains. As visible from Fig.~\ref{fgr:2}(c) and \ref{fgr:3}(a), this is not true for all initially transported SPBs; some get immobilized at larger or lower DW separation distances, and a few even don't get immobilized at all. This reflects the spread of magnetophoretic mobility within a single badge of nominally equal-sized particles, as can also be deduced from the SPB fractionation experiment of Fig.~\ref{fgr:5}. Hence, our MFL design is demonstrated to enable sorting of SPBs by their mobility, so that subsequent bio-detection steps are inducible using SPBs of a defined mobility. For the practical realization of this sorting routine, the experiments have shed light on the impact of interparticle interactions. It was observed that in most cases, not all initially mobile SPBs could be transported back into the opposite direction after immobilization (see Fig.~\ref{fgr:3}(a)). The reason for this can be traced back to the interaction of an immobilized SPB with an incoming, movable SPB: The incoming bead may push the immobilized bead to an adjacent DW position with an even larger domain stripe width before the incoming bead itself eventually gets immobilized. As a result, the pushed bead may not be able to travel back into the opposite direction, as its transport step distance into the opposite direction is now larger than the transport step distance before the collision with the incoming bead. Conversely, SPBs may regain their mobility due to collision with approaching SPBs, which is occasionally observable in the intensity map of Fig.~\ref{fgr:3}(a). This collision effect pronounces that the SPB concentration is another important factor for the fractionation efficiency when using specifically designed MFLs. This was verified by experimenting with a dramatically increased SPB concentration, where a large fraction of mobile beads was formed when using the same parameters for the external magnetic field sequence.
\section{Conclusion}
Parallel-stripe domains of gradually varied stripe width were imprinted into an EB thin film substrate to engineer a magnetic stray field landscape for the controlled spatial fractionation of liquid-dispersed SPBs transported close to the substrate surface. Ion bombardment induced magnetic patterning was employed to fabricate a hh/tt alternating magnetization configuration within the stripe domains, with the stripe width being periodically increased and decreased between \SI{1.2}{\micro\meter} and \SI{10}{\micro\meter}. The resulting MFL on top of the substrate leads to the capture and a stepwise, unidirectional transport of water-dispersed SPBs upon external application of a sequence of trapezoidal magnetic field pulses. SPBs with a diameter of \SI{2.8}{\micro\meter} were observed using an optical brightfield microscope to assemble in vertical rows on top of DW positions in the underlying substrate. Thus, the horizontal distance between adjacent SPB rows is locally different due to the gradually varied DW separation distance. Accordingly, the DW separation distance defines the transport step distance for one particular SPB row. Reaching the location of largest stripe domain width, SPBs transitioned to an oscillatory motion for a given configuration of external field pulses, resulting in their effective immobilization. Therefore, the SPB motion behavior is a function of lateral position, presuming a fixed frequency for the external field sequence. This is a significant advancement compared to established works, where the frequency of the external time-varying field needs to be tuned to initiate a transition from mobile to immobile SPBs. The locally different SPB motion regimes for our MFL design can be attributed to the lateral change in the transport step distance itself, as well as an observed decrease in SPB steady-state motion velocity with increasing transport step distance. While observing a spatial fractionation of nominally equal-sized SPBs due to statistical variations in their magnetophoretic mobility, a proof-of-concept experiment was finally conducted to deterministically fractionize SPBs of significantly different sizes. For this, a mixture of SPBs with diameters of \SI{1.0}{\micro\meter} and \SI{2.8}{\micro\meter} was placed on top of the substrate and all SPBs were transported to the largest stripe domains via an external field sequence of fixed frequency. After merely \SI{3}{\second}, all SPBs performed oscillatory motion but with a clear spatial distinction between \SI{1.0}{\micro\meter}- and \SI{2.8}{\micro\meter}-sized SPBs. Smaller SPBs were, on average, immobilized earlier (at smaller DW separation distance) than larger SPBs, leading to a significant spatial fractionation of both species. Since the pattern of sequentially increasing/decreasing stripe domain width is periodically repeated throughout the whole substrate, the fractionation is parallelized for all on-chip present SPBs. This result may pave the way for high-throughput and time-efficient fractionation of magnetic particle species as a function of properties that are more difficult to disentangle, for instance, their surface characteristics. Since particles are transported in the vicinity of the substrate surface for our investigated system, changes in the surface properties lead to variations in the substrate surface-particle surface interaction. This, in turn, influences the separation distance between the two surfaces with an expected impact on the particle magnetophoretic mobility, e.g., via the modified friction situation. Optimizing the MFL design for this functionality is one of our future tasks, rendering purpose-oriented MFLs as promising tools for the practical implementation of magnetic particle sorting routines in LOC devices.

\section*{Acknowledgments}
RH, LP, and AE acknowledge project and scientific infrastructure funding by the German Research Foundation (DFG) under the project numbers 514858524, 433501699, 361379292, and 361396165. PK acknowledges financial support from the National Science Centre, Poland, under grant number 2020/39/B/ST5/01915. The authors acknowledge fruitful discussions with M. Vogel, Kiel University.

\section*{Author Contributions}
RH: Conceptualization, Investigation, Formal Analysis, Visualization, Writing - Original Draft\\
LP: Investigation, Formal Analysis, Writing - Review \& Editing\\
PK: Investigation, Resources, Writing - Review \& Editing\\
AE: Conceptualization, Resources, Supervision, Project Administration, Funding Acquisition, Writing - Review \& Editing
\bibliographystyle{unsrt}  
\bibliography{library}  

\begin{thebibliography}{10}

\bibitem{Kricka2001}
Larry~J Kricka.
\newblock Microchips, microarrays, biochips and nanochips: personal laboratories for the 21st century.
\newblock {\em Clinica Chimica Acta}, 307:219--223, 5 2001.

\bibitem{Knight2002}
Jonathan Knight.
\newblock Honey, i shrunk the lab.
\newblock {\em Nature}, 418:474--475, 8 2002.

\bibitem{Manz1990}
A.~Manz, N.~Graber, and H.~M. Widmer.
\newblock Miniaturized total chemical analysis systems: A novel concept for chemical sensing.
\newblock {\em Sensors and Actuators: B. Chemical}, 1:244--248, 1 1990.

\bibitem{Yetisen2013}
Ali~Kemal Yetisen, Muhammad~Safwan Akram, and Christopher~R. Lowe.
\newblock Paper-based microfluidic point-of-care diagnostic devices.
\newblock {\em Lab on a Chip}, 13:2210--2251, 5 2013.

\bibitem{Cardoso2017}
S~Cardoso, D~C Leitao, T~M Dias, J~Valadeiro, M~D Silva, A~Chicharo, V~Silverio, J~Gaspar, and P~P Freitas.
\newblock Challenges and trends in magnetic sensor integration with microfluidics for biomedical applications.
\newblock {\em Journal of Physics D: Applied Physics}, 50:213001, 6 2017.

\bibitem{Zhang2020}
Yi~Zhang, Xiao Liu, Lingling Wang, Hanjie Yang, Xiaoxiao Zhang, Chenglong Zhu, Wenlong Wang, Lijing Yan, and Bowei Li.
\newblock Improvement in detection limit for lateral flow assay of biomacromolecules by test-zone pre-enrichment.
\newblock {\em Scientific reports}, 10:9604, 6 2020.

\bibitem{Pankhurst2003}
Q~A Pankhurst, J~Connolly, S~K Jones, and J~Dobson.
\newblock Applications of magnetic nanoparticles in biomedicine.
\newblock {\em Journal of Physics D: Applied Physics}, 36:R167--R181, 7 2003.

\bibitem{Gijs2004}
Martin A.~M. Gijs.
\newblock Magnetic bead handling on-chip: New opportunities for analytical applications.
\newblock {\em Microfluidics and Nanofluidics}, 1:22--40, 10 2004.

\bibitem{Pamme2006}
Nicole Pamme.
\newblock Magnetism and microfluidics.
\newblock {\em Lab on a Chip}, 6:24--38, 12 2006.

\bibitem{Ruffert2016}
Christine Ruffert.
\newblock Magnetic bead—magic bullet.
\newblock {\em Micromachines}, 7:21, 1 2016.

\bibitem{Gao2013}
Yang Gao, Alexander~Van Reenen, Martien~A. Hulsen, Arthur M.~De Jong, Menno~W.J. Prins, and Jaap M.J.~Den Toonder.
\newblock Disaggregation of microparticle clusters by induced magnetic dipole-dipole repulsion near a surface.
\newblock {\em Lab on a Chip}, 13:1394--1401, 3 2013.

\bibitem{Moerland2019}
C.~P. Moerland, L.~J.~Van IJzendoorn, and M.~W.J. Prins.
\newblock Rotating magnetic particles for lab-on-chip applications-a comprehensive review.
\newblock {\em Lab on a Chip}, 19:919--933, 3 2019.

\bibitem{Ran2014}
Ying~Fen Ran, Conor Fields, Julien Muzard, Viktoryia Liauchuk, Michael Carr, William Hall, and Gil~U. Lee.
\newblock Rapid, highly sensitive detection of herpes simplex virus-1 using multiple antigenic peptide-coated superparamagnetic beads.
\newblock {\em Analyst}, 139:6126--6134, 2014.

\bibitem{Rampini2021}
Stefano Rampini, Peng Li, Dhruv Gandhi, Marina Mutas, Ying~Fen Ran, Michael Carr, and Gil~U. Lee.
\newblock Design of micromagnetic arrays for on-chip separation of superparamagnetic bead aggregates and detection of a model protein and double-stranded dna analytes.
\newblock {\em Scientific Reports}, 11:5302, 12 2021.

\bibitem{Reginka2021}
Meike Reginka, Hai Hoang, Özge Efendi, Maximilian Merkel, Rico Huhnstock, Dennis Holzinger, Kristina Dingel, Bernhard Sick, Daniela Bertinetti, Friedrich~W. Herberg, and Arno Ehresmann.
\newblock Transport efficiency of biofunctionalized magnetic particles tailored by surfactant concentration.
\newblock {\em Langmuir}, 37:8498--8507, 7 2021.

\bibitem{Feely2023}
Nathan Feely, Anita Wdowicz, Anne Chevalier, Ying Wang, Peng Li, Fanny Rollo, and Gil~U. Lee.
\newblock Targeting mucin protein enables rapid and efficient ovarian cancer cell capture: Role of nanoparticle properties in efficient capture and culture.
\newblock {\em Small}, 19:2207154, 5 2023.

\bibitem{Wise2015}
Naomi Wise, David~Tim Grob, Karl Morten, Ian Thompson, and Steve Sheard.
\newblock Magnetophoretic velocities of superparamagnetic particles, agglomerates and complexes.
\newblock {\em Journal of Magnetism and Magnetic Materials}, 384:328--334, 6 2015.

\bibitem{Zhou2016}
Chen Zhou, Eugene~D. Boland, Paul~W. Todd, and Thomas~R. Hanley.
\newblock Magnetic particle characterization—magnetophoretic mobility and particle size.
\newblock {\em Cytometry Part A}, 89:585--593, 6 2016.

\bibitem{Donolato2012}
Marco Donolato, Bjarke~Thomas Dalslet, and Mikkel~Fougt Hansen.
\newblock Microstripes for transport and separation of magnetic particles.
\newblock {\em Biomicrofluidics}, 6:024110, 6 2012.

\bibitem{Rampini2016}
S.~Rampini, P.~Li, and G.~U. Lee.
\newblock Micromagnet arrays enable precise manipulation of individual biological analyte–superparamagnetic bead complexes for separation and sensing.
\newblock {\em Lab on a Chip}, 16:3645--3663, 9 2016.

\bibitem{Block2023}
Findan Block, Finn Klingbeil, Umer Sajjad, Christine Arndt, Sandra Sindt, Dennis Seidler, Lars Thormählen, Christine Selhuber‐Unkel, and Jeffrey McCord.
\newblock Magnetic bucket brigade transport networks for cell transport.
\newblock {\em Advanced Materials Technologies}, 4 2023.

\bibitem{Tierno2009}
Pietro Tierno, Francesc Sagués, Tom~H. Johansen, and Thomas~M. Fischer.
\newblock Colloidal transport on magnetic garnet films.
\newblock {\em Physical Chemistry Chemical Physics}, 11:9615--9625, 10 2009.

\bibitem{Holzinger2015a}
Dennis Holzinger, Iris Koch, Stefan Burgard, and Arno Ehresmann.
\newblock Directed magnetic particle transport above artificial magnetic domains due to dynamic magnetic potential energy landscape transformation.
\newblock {\em ACS Nano}, 9:7323--7331, 7 2015.

\bibitem{Urbaniak2024}
Maciej Urbaniak, Daniel Kiphart, Michał Matczak, Feliks Stobiecki, Gabriel~David Chaves-O’Flynn, and Piotr Kuświk.
\newblock Ferrimagnetic tb/co multilayers patterned by ion bombardment as substrates for magnetophoresis.
\newblock {\em Scientific Reports}, 14:23771, 2024.

\bibitem{Yellen2007}
Benjamin~B. Yellen, Randall~M. Erb, Hui~S. Son, Rodward Hewlin, Hao Shang, and Gil~U. Lee.
\newblock Traveling wave magnetophoresis for high resolution chip based separations.
\newblock {\em Lab on a Chip}, 7:1681--1688, 11 2007.

\bibitem{Yellen2009}
Benjamin~B. Yellen and Lawrence~N. Virgin.
\newblock Nonlinear dynamics of superparamagnetic beads in a traveling magnetic-field wave.
\newblock {\em Physical Review E}, 80:011402, 2009.

\bibitem{Huhnstock2024}
Rico Huhnstock, Lukas Paetzold, Maximilian Merkel, Piotr Kuświk, and Arno Ehresmann.
\newblock Combined funnel, concentrator, and particle valve functional element for magnetophoretic bead transport based on engineered magnetic domain patterns.
\newblock {\em Small}, 20:2305675, 3 2024.

\bibitem{Kim2025}
Hyeonseol Kim, Abbas Ali, Yumin Kang, Byeonghwa Lim, and CheolGi Kim.
\newblock Surface-driven particle dynamics: Sequential synchronization of colloidal flow attempted in a static fluidic environment.
\newblock {\em ACS Applied Materials \& Interfaces}, 17:12772--12781, 2025.

\bibitem{Abedini-Nassab2025}
Roozbeh Abedini-Nassab, Milad~Alishahi Toosi, Yinchu Shen, Fatemeh Maghsoodi, and Yaping Dan.
\newblock Magnetophoretic capacitors operating in a tri-axial magnetic field for on-chip bioapplications.
\newblock {\em IEEE Sensors Journal}, 25:382--389, 2025.

\bibitem{Liu2007}
Chengxun Liu, Liesbet Lagae, Roel Wirix-Speetjens, and Gustaaf Borghs.
\newblock On-chip separation of magnetic particles with different magnetophoretic mobilities.
\newblock {\em Journal of Applied Physics}, 101:024913, 1 2007.

\bibitem{Mougin2001}
A.~Mougin, S.~Poppe, J.~Fassbender, B.~Hillebrands, G.~Faini, U.~Ebels, M.~Jung, D.~Engel, A.~Ehresmann, and H.~Schmoranzer.
\newblock Magnetic micropatterning of feni/femn exchange bias bilayers by ion irradiation.
\newblock {\em Journal of Applied Physics}, 89:6606--6608, 2001.

\bibitem{Ehresmann2004}
A~Ehresmann, I~Krug, A~Kronenberger, A~Ehlers, and D~Engel.
\newblock In-plane magnetic pattern separation in nife/nio and co/nio exchange biased bilayers investigated by magnetic force microscopy.
\newblock {\em Journal of Magnetism and Magnetic Materials}, 280:369--376, 9 2004.

\bibitem{Ehresmann2015}
Arno Ehresmann, Iris Koch, and Dennis Holzinger.
\newblock Manipulation of superparamagnetic beads on patterned exchange-bias layer systems for biosensing applications.
\newblock {\em Sensors (Switzerland)}, 15:28854--28888, 11 2015.

\bibitem{Lengemann2012}
Daniel Lengemann, Dieter Engel, and Arno Ehresmann.
\newblock Plasma ion source for in situ ion bombardment in a soft x-ray magnetic scattering diffractometer.
\newblock {\em Review of Scientific Instruments}, 83:53303, 2012.

\bibitem{Grob2018}
David~Tim Grob, Naomi Wise, Olayinka Oduwole, and Steve Sheard.
\newblock Magnetic susceptibility characterisation of superparamagnetic microspheres.
\newblock {\em Journal of Magnetism and Magnetic Materials}, 452:134--140, 2018.

\bibitem{Vansteenkiste2014}
Arne Vansteenkiste, Jonathan Leliaert, Mykola Dvornik, Mathias Helsen, Felipe Garcia-Sanchez, and Bartel~Van Waeyenberge.
\newblock The design and verification of mumax3.
\newblock {\em AIP Advances}, 4:107133, 10 2014.

\bibitem{Wirix-Speetjens2005}
Roel Wirix-Speetjens, Wim Fyen, Kaidong Xu, Jo~De Boeck, and Gustaaf Borghs.
\newblock A force study of on-chip magnetic particle transport based on tapered conductors.
\newblock {\em IEEE Transactions on Magnetics}, 41:4128--4133, 10 2005.

\end{thebibliography}
\newpage
\renewcommand\thefigure{S\arabic{figure}} 
\section*{Supplementary Information}
\setcounter{figure}{0}
\subsection*{Resist structure utilized for magnetic patterning}
\begin{figure*}[ht!]
 \centering
 \includegraphics[width=\linewidth]{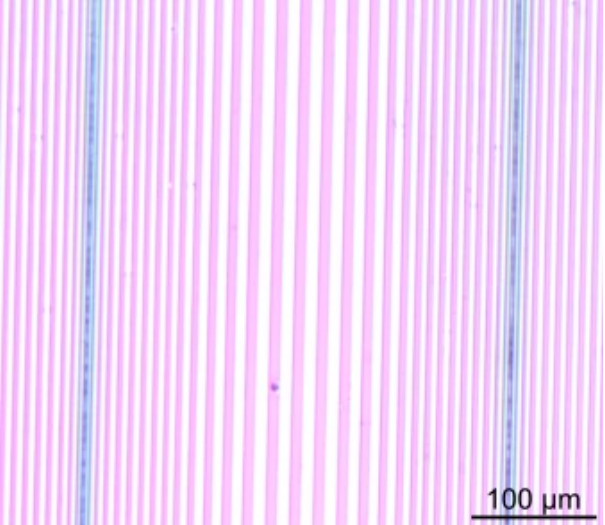}
 \caption{Image of lithographically structured resist utilized for the ion bombardment induced magnetic patterning of an exchange-biased thin film system, obtained by optical microscopy.}
 \label{fgr:S1}
\end{figure*}
\newpage
\subsection*{Velocity during a transport step as a function of step distance for all observed particles}
\begin{figure*}[ht!]
 \centering
 \includegraphics[width=\linewidth]{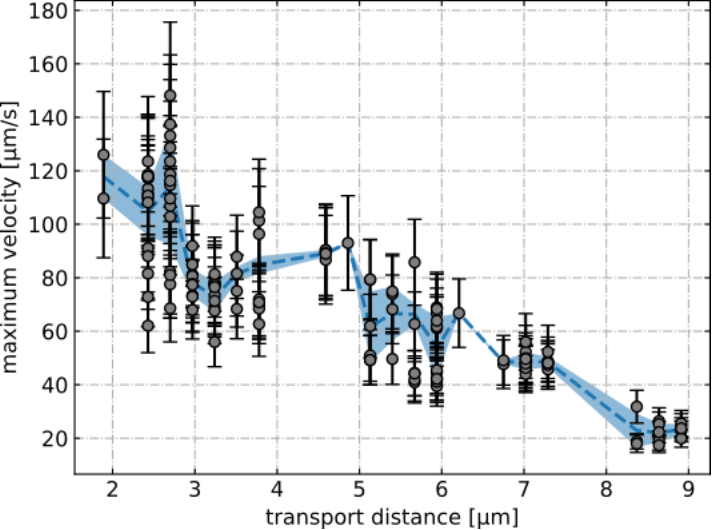}
 \caption{Maximum step velocity as a function of transport step distance determined for micron-sized superparamagnetic beads moving on top of a magnetically stripe domain patterned substrate with gradually increasing stripe width. Error bars indicate the uncertainty of velocity for a single transport step. The blue dashed line marks the development of the average maximum velocity with an overall decreasing tendency with increasing transport step distance/stripe domain width. The blue shaded area indicates the standard deviation of the average velocities.}
 \label{fgr:S2}
\end{figure*}
\subsection*{Simulation of magnetic stray field landscape}
The simulation package MuMax3 [1] was used to compute the magnetization distribution $\vec{m}(x,y)$ within a region of interest for the investigated stripe domain pattern with gradually increasing stripe width. Depending on whether stripe regions were bombarded/non-bombarded during the fabrication procedure, differing magnetic properties were assigned: An exchange stiffness constant of $A_{\mathrm{ex}}$ = \SI{3e-11}{\joule\per\meter} [2], a saturation magnetization of $M_{\mathrm{S}}$ = \SI{1.23e6}{\ampere\per\meter} [3], a uniaxial anisotropy constant of $K$ = \SI{4.5e4}{\joule\per\meter\cubed} [4] for the non-bombarded stripes, and accordingly $A_{\mathrm{ex}}$ = \SI{3e-11}{\joule\per\meter}, $M_{\mathrm{S}}$ = \SI{1.18e6}{\ampere\per\meter}, $K$ = \SI{3.375e4}{\joule\per\meter\cubed} for the bombarded stripes. The saturation magnetization and anisotropy constant are slightly reduced for the bombarded regions, as reported in previous publications [3,5]. For implementing the exchange bias-related pinning of the respective domain magnetizations, additional biasing magnetic fields were defined for bombarded/non-bombarded stripes with opposing directions. Here, the magnetic flux densities were chosen to be \SI{13}{\milli\tesla} for the non-bombarded regions and \SI{6.7}{\milli\tesla} for the bombarded stripes according to experimentally determined values from hysteresis loop measurements. The region of interest was discretized into cubic elements of \SI{5}{\nano\meter} × \SI{5}{\nano\meter} × \SI{10}{\nano\meter}. The simulation software computed the relaxed magnetization state of the described system, which was subsequently used for obtaining magnetic stray field components $\vec{H}_z(\vec{r})$ and $\vec{H}_x(\vec{r})$ (see Figures S3 and S4) via the following dipole approximation [6]:
\begin{equation*}
    \vec{H}(\vec{r})=\frac{1}{4\pi}\cdot\sum_{i}\frac{3\vec{R}(\vec{R}\cdot\vec{m}_{i})}{\left|\vec{R}\right|^{5}}-\frac{\vec{m}_{i}}{\left|\vec{R}\right|^{3}}.
\label{eq:S}
\end{equation*}
$\vec{R} = \vec{r}-\vec{r}_i$ represents the distance vector between position $\vec{r}$ and dipole position (mesh element position) $\vec{r}_i$.
\begin{figure*}[ht!]
 \centering
 \includegraphics[width=\linewidth]{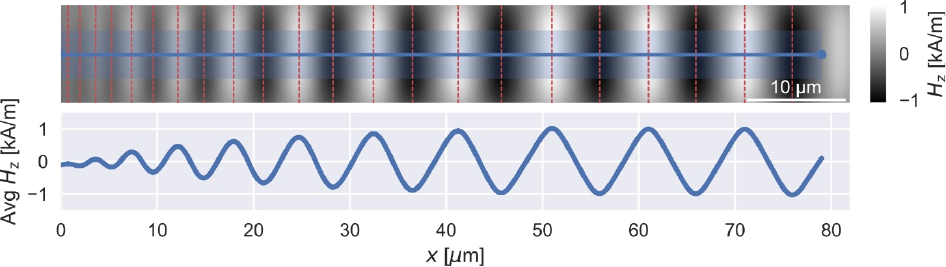}
 \caption{Spatial distribution of the magnetic stray field component $H_z$ emerging from an exchange-biased thin film system patterned with magnetic parallel stripe domains with gradually increasing stripe width. Red dashed lines indicate boundaries of stripe domains (domain walls). The field distribution was derived from micromagnetic simulations for a distance of \SI{2500}{\nano\meter} above the magnetic thin film system. A line profile visualizes the lateral evolution of $H_z$, averaged over the blue-shaded area.}
 \label{fgr:S3}
\end{figure*}
\begin{figure*}[ht!]
 \centering
 \includegraphics[width=\linewidth]{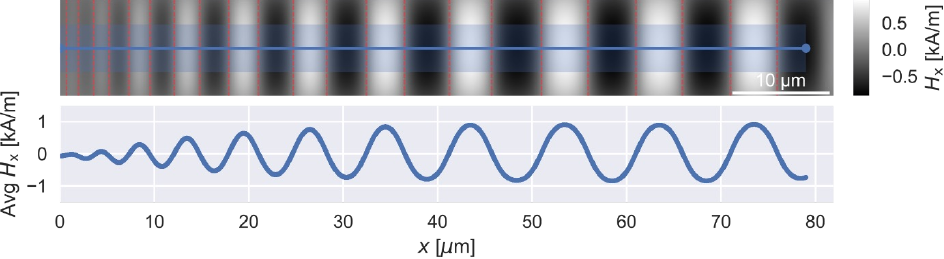}
 \caption{Spatial distribution of the magnetic stray field component $H_x$ emerging from an exchange-biased thin film system patterned with magnetic parallel stripe domains with gradually increasing stripe width. Red dashed lines indicate boundaries of stripe domains (domain walls). The field distribution was derived from micromagnetic simulations for a distance of \SI{2500}{\nano\meter} above the magnetic thin film system. A line profile visualizes the lateral evolution of $H_x$, averaged over the blue-shaded area.}
 \label{fgr:S4}
\end{figure*}
\subsection*{References for Supplementary Information}
[1] A. Vansteenkiste, J. Leliaert, M. Dvornik, M. Helsen, F. Garcia-Sanchez, B. Van Waeyenberge, AIP Advances, 2014, 4, 107133.

[2] D. V. Berkov, C. T. Boone, I. N. Krivorotov, Physical Review B - Condensed Matter and Materials Physics, 2011, 83, 54420.

[3] H. Huckfeldt, A. Gaul, N. D. Müglich, D. Holzinger, D. Nissen, M. Albrecht, D. Emmrich, A. Beyer, A. Gölzhäuser, A. Ehresmann, Journal of Physics: Condensed Matter, 2017, 29, 125801.

[4] D. Holzinger, N. Zingsem, I. Koch, A. Gaul, M. Fohler, C. Schmidt, A. Ehresmann, Journal of Applied Physics, 2013, 114, 013908.

[5] N. D. Müglich, M. Merkel, A. Gaul, M. Meyl, G. Götz, G. Reiss, T. Kuschel, A. Ehresmann, New Journal of Physics, 2018, 20, 053018.

[6] Nolting, W. Grundkurs Theoretische Physik 3. Springer (2013).
\end{document}